\documentclass[a4paper,11pt]{article}
\usepackage{pos}

\usepackage{epsf}

\def\NPB#1#2#3{{\it Nucl.\ Phys.}\/ {\bf B#1} (#2) #3}
\def\PLB#1#2#3{{\it Phys.\ Lett.}\/ {\bf B#1} (#2) #3}

\def\PRD#1#2#3{{\it Phys.\ Rev.}\/ {\bf D#1}  (#2) #3}
\def\PRL#1#2#3{{\it Phys.\ Rev.\ Lett.}\/ {\bf #1} (#2) #3}
\def\PRT#1#2#3{ {\it Phys.\ Rep.}\/ {\bf#1} (#2) #3}
\def\MODA#1#2#3{ {\it Mod.\ Phys.\ Lett.}\/ {\bf A#1} (#2) #3}

\def\IJMP#1#2#3{ {\it Int.\ J.\ Mod.\ Phys.}\/ {\bf A#1} (#2) #3}

\def\APP#1#2#3{ {\it Astropart.\ Phys.}\/ {\bf #1} (#2) #3}
\def\EJP#1#2#3{ {\it Eur.\ Phys.\ Jour.}\/ {\bf C#1} (#2) #3}
\def\JHEP#1#2#3{ {\it JHEP}\/ {\bf #1} (#2) #3}

\def\etal{{\it et al\/}}

\def\AEF{A.E. Faraggi}

\newcommand{\beq}{\begin{equation}}
\newcommand{\eeq}{\end{equation}}
\newcommand{\beqa}{\begin{eqnarray}}
\newcommand{\beqn}{\begin{eqnarray}}
\newcommand{\eeqn}{\end{eqnarray}}
\newcommand{\eeqa}{\end{eqnarray}}

\newcommand{\ba}{\begin{eqnarray}}
\newcommand{\ea}{\end{eqnarray}}

\title{Fundamental or Composite? The Higgs Enigma}

\author*[a]{Alon E Faraggi}

\affiliation[a]{
  Mathematical Sciences Department, University of Liverpool,
  Liverpool L69 7ZL, United Kingdom}

\emailAdd{alon.faraggi@liv.ac.uk}

\abstract{
  The discovery of the Higgs boson by the ATLAS and CMS experiments
  concluded a glorious century of experimental particle physics
  discoveries, from Rutherford's discovery of the nucleus in
  1911, through the discoveries of quarks and leptons from the
  1950s to the 1970s, to the discoveries of the weak vector bosons
  in the 1980s. It cemented the Standard Model of particle physics
  as providing the viable parameterisation of all sub--atomic observables up
  to the TeV scale and possibly up to the GUT and Planck scales.
  The experimental determination of the Higgs properties and
  parameters will shed light on these fundamental theories. A particularly
  pertaining question from the point of view of String Phenomenology is
  whether the Higgs boson is a fundamental or composite particle.
  The fermionic $Z_2\times Z_2$ orbifolds provide bench mark models to
  explore how the parameters of the Standard Model can arise from a
  theory of quantum gravity, as well as for physics
  Beyond the Standard Model. 
  Observation that the Higgs is composite will nullify much of the work
  that have gone into heterotic--string model building over the past
  $\sim$40 years and will indicate the relevance of other classes of
  string vacua or possibly other approaches to quantum gravity.
  An ideal facility in the near future to investigate this question is a
  hadron collider at 50-–60 TeV that utilises contemporary magnet
  technology and can be built in 10–15 years from decision.
}

\FullConference{Proceedings of the Corfu Summer Institute
  2025 "School and Workshops on Elementary Particle Physics and Gravity"
  (CORFU2025)\\
27 April - 28 September, 2025\\
Corfu, Greece\\}

\begin{document}
\maketitle

\section{Introduction}

The discovery of the Higgs boson by the ATLAS \cite{ATLAS:2012yve}
and CMS \cite{CMS:2012qbp} experiments in July 2012,
concluded a century of discoveries that cemented the
Standard Model (SM) of particle physics as providing the
mathematical parameterisation of all observable sub--atomic
data to date. A celebration of human curiosity and ingenuity
to uncover the most fundamental layers of reality.
One cannot give high enough praise 
to the experimental physicists on the front line.
The heroes on the subatomic frontier in a relentless
effort to go beyond where no one has gone before.
The confirmation of the 
of the Standard Model by the experimental observations
as the viable parameterisation opens the door to
questions about its fundamental origins.

The Standard Model is composed of the gauge, matter and Higgs sectors,
governed by nineteen parameters. Inclusion of massive neutrinos
increases the number of parameters to 26 or 28 if neutrinos are Majorana
fermions. The gauge sector consists of the $SU(3)\times SU(2)_L\times U(1)_Y$
gauge group factors, introducing the mediators of the gauge group interactions
that transform in the adjoint representations of the different groups.
The gauge coupling parameters determine the strength of the interactions
with the scalar, fermion and gauge content of the model and any additional
states, Beyond the Standard Model (BSM) that are charged under the Standard
Model group factors. It is notable that at lower energies, the symmetries
of the SM are not manifest, but only a $SU(3)\times U(1)_{e.m.}$ remnant.
The $SU(2)_L\times U(1)_Y$ symmetry is broken by the Higgs mechanism
to $U(1)_{e.m.}$ and explains how the massive vector bosons of the
broken weak interaction can co--exist with the massless vector boson
of the electromagnetic interaction. While the theoretical explanation of
this mechanism was proposed in the mid--1960s, its experimental confirmation
had to await nearly fifty years.

The fermionic matter sector of the SM consists of three families of
quarks and leptons in representations of the SM group factors.
Only the left--handed fields interact via the weak interactions.
Adding a SM singlet as a right--handed neutrino, the small mass of the
left--handed neutrinos is explained by the seesaw mechanism that gives
large Majorana mass to the right--handed neutrino, whereas the Dirac
mass term is governed by the Vacuum Expectation Value (VEV) of the Higgs
field, which is of the order of the electroweak scale.
Each SM model family has sixteen states, whereas the scalar sector
of the SM consists of a single electroweak doublet field. The masses
of the SM fields are generated by their couplings to the Higgs field.
These can be written in terms of the six quark and three charged
lepton masses; the three angles and on CP--violating phase in
the Cabibbo--Kobayashi--Maskawa mixing matrix; the Higgs mass and
the Higgs VEV. An additional undetermined parameter is a parameter that
measures the strength of CP--violation in the strong interactions. The
smallness of this parameter is one of the ad hoc features of the SM that
motivates its speculative extensions. The extension of the SM to include
massive neutrinos increases the number of parameters to twenty--six,
being the three neutrino masses and the parameters in the
Pontecorvo–Maki–Nakagawa–Sakata matrix. 

The Standard model parametrises numerous experimental observables in terms
of these 26 parameters \cite{PDG2024}. Some of these observables are
measured to a very high degree of accuracy, {\it e.g.} the magnetic moment
of the electron. While the theoretical structure of the SM was set
by the early 70s, the experimental validation of its different
components and their relations took further four decades, culminating
with the discovery of the Higgs particle in 2012. The four LEP experiments
measured the parameters of the weak interactions to an unprecedented accuracy,
including their non--Abelian character. The logarithmic evolution of the
SM parameters was also confirmed experimentally, promoting the
possibility that the SM continues to provide a viable perturbative
parametrisation of all observable data up to energy scales that exceeds by
far those that are accessible in contemporary experiments.
Moreover, the measured values of the couplings of the SM gauge group factors
are compatible with the notion that the three SM gauge groups unify at a
very high energy scale. This feat is only compatible with the data
in the presence of three generations. Remarkably, three replicas of
of the Standard Model generations is precisely what is observed in
particle physics experiments. Furthermore, the Standard Cosmological
Model favours the existence of three left--handed neutrinos, in
agreement with the existence of three and only three chiral generations. 
Further indication for the
high scale unification arise from the longevity of the
proton and suppression of the left--handed neutrino masses.

Given the experimental data from terrestrial, astrophysical and cosmological
observations that all agree with the Standard Particle Model, the most
prudent assumption is to assume that the Standard Model is all there is.
The are several provisos to this stipulation.
The first is that there are at least two parameters in the SM
that need to be fine tuned. The first is the mass of the Higgs boson.
Its mass is not protected by a symmetry and is affected by any
new physics at a higher scale that can vary from the $\sim$TeV scale
to the Planck scale. The corrections to the Higgs mass are
quadratic in the cutoff scale which entails a high degree of
fine tuning if the cutoff scale is high. The parameter
that parametrises CP--violation in the strong interactions
is also fine tuned to a high degree in the Standard Model. 

The second major proviso to the success of the SM
is that gravity is not included in the SM.
There are several basic conflicts
between gravitational based observations and the SM.
The first is the need for additional matter
to account for galactic and cosmological data. The
second and more fundamental is the disagreement between
the gravitationally based inferences of the vacuum
energy versus the vacuum energy predicted in the
SM, and, more generally, in Quantum Field Theories.
This dichotomy can only be resolved by developing
a unified formalism for gravity and quantum mechanics.
It should be remarked that none of the contemporary
approaches to quantum gravity can resolve the vacuum
energy problem. 

\begin{figure}[h]
\includegraphics[width=18pc]{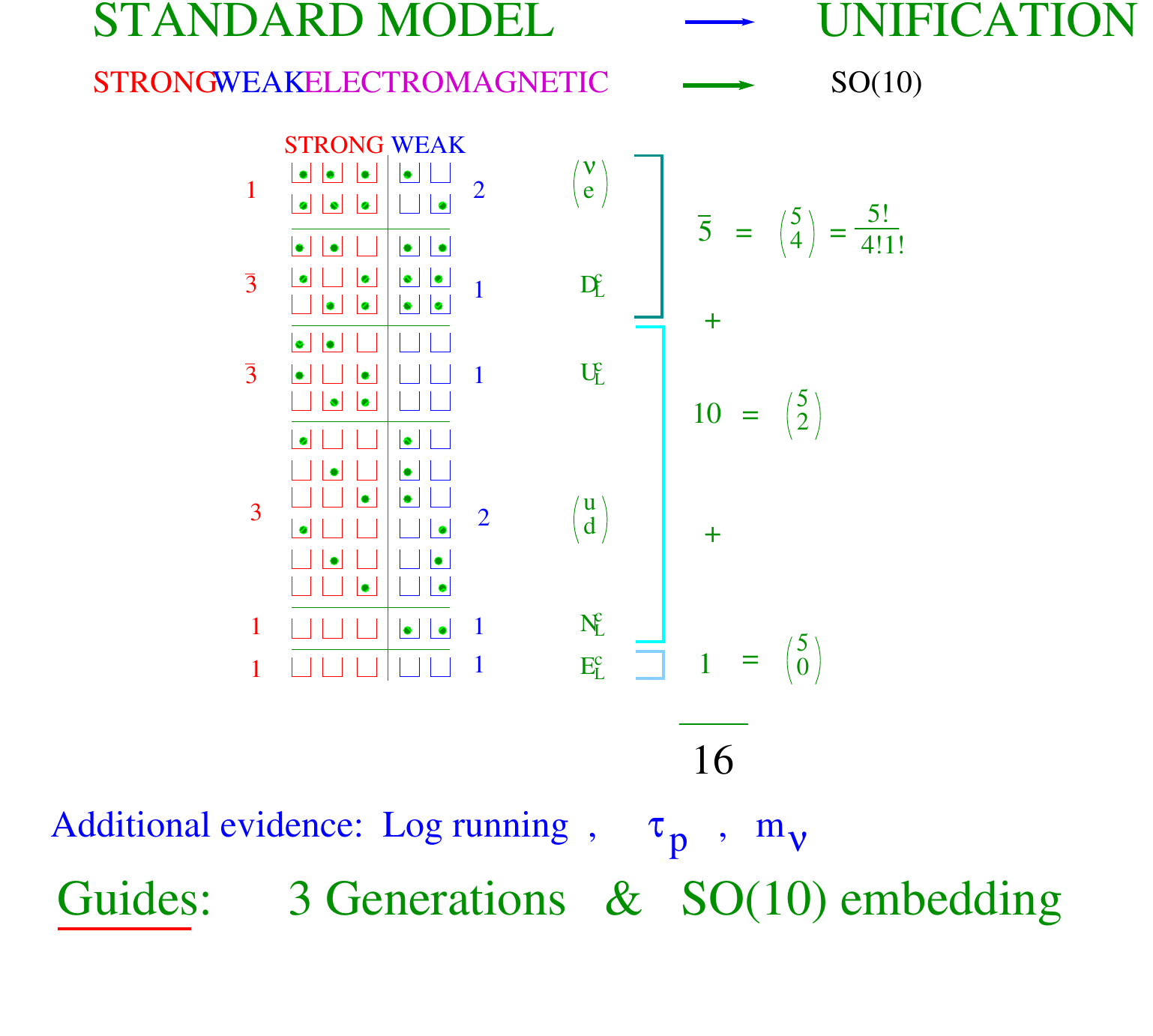}\hspace{2pc}%
\begin{minipage}[b]{10pc}
\caption{\label{figure1}%
\emph{
The Standard Model (SM) gauge and matter spectrum favours
its incorporation in Grand Unified Theories (GUTs). 
In $SO(10)$ GUT each generation of the Standard Model
fits into a single spinorial
$16$ representation. Further  
evidence for high--scale unification arises from:
proton longevity;
suppression of left--handed neutrino masses;
logarithmic evolution of the SM parameters.
}
}
\end{minipage}
\end{figure}

What is the lesson to extract from the success of the SM. 
Perhaps the most remarkable feature of the SM is the
fact that its matter states fit in representations of
larger gauge groups. Most compelling is the embedding
of the Standard Model in $SO(10)$ Grant Unified Theory,
in which each of the Standard Model generations, plus the
right--handed neutrino, fits in a single spinorial $16$
representation of $SO(10)$. This remarkable coincidence is
depicted in figure \ref{figure1}.
Given that the SM spectrum was discovered experimentally, rather
than imposed by mathematical consistency, this embedding correlates
54 observable parameters that include the number of multiplets
per family, times the number of families, times the number
of group factors. It is important to note that the number of
families is restricted by the experimental data to be three and
only three chiral generations, whereas additional heavy non--chiral
families are allowed, albeit restricted, by the data.

Although the embedding of the SM in Grand Unified Theories,
and in particular $SO(10)$ GUT is well motivated by the
experimental data, it is clearly not the end of the road.
In particular, Grand Unified Theories do not account for the
multiplicity of families nor for the fermion masses and mixing
parameters. To seek a calculational framework for these parameters,
gravity has to be brought into the fold. String theory is a mundane
extension of the concept of particles with well defined
physical attributes. In point particle Quantum Field Theories (QFTs),
we parametrise the path of elementary particles as idealised points.
It is important to note that the particles are not actual point particles,
but rather fields spreading all over spacetime, but the interactions
between particles are generated at points in spacetime, which leads
to singularities and divergent amplitudes.
In the SM the divergences are absorbed into a
finite number of experimentally measured parameters, and
only renormalisable parameters are incorporated in the SM.
As we expect the SM model to be valid only up to some cutoff scale,
non--renormalisable operators will be induced at that scale,
including operators that violate baryon and lepton number and
can induce rapid proton decay. The proton lifetime, which exceeds
$10^{34}$ years, suggests that the cutoff scale has to be
very high, possibly at the Planck scale, where the gravitational
interaction becomes relevant.
In string theory the singularities are smoothed out leading to
a perturbatively finite theory of quantum gravity.
are spread out which smoothies the singularities.
String theory 
provides a self consistent framework to explore the
synthesis of the SM and gravity. 

\section{Embedding the Standard Model in string theory}

It is often said that string theory gives rise to $10^{500}$ vacua
and therefore lacks predictive power. This is of course a questionable argument.
In that sense quantum field theories do not have predictive power either.
The particle content of the SM is not determined by any consistency
criteria but was deduced from countless experimental observations.
Furthermore, the parameters in the SM are continuous variables and
give rise to an infinite number of possibilities.
String phenomenology aims to connect between string theory and
observational data by constructing and analysing string vacua
that reproduce some basic phenomenological characteristics
of the SM. Among them, the existence of three chiral generations
and their embedding in $SO(10)$ spinorial representations. There is a
relatively small number of string theories in ten dimensions
that are all supposed to be connected by perturbative and
non--perturbative duality transformations. 
However, only the $E_8\times E_8$ heterotic--string gives rise
to spinorial representations in its perturbative massless spectrum.
Phenomenological string models can therefore be
explored by studying compactifications of the $E_8\times E_8$
heterotic--string. Whether these phenomenological string models
prove to be relevant for our actually observed world is, of course,
desired, but is of secondary importance. The primary aim is to develop the
tools to connect between string theory and observational data
in its varied forms. One of those forms is the data encoded in the
Standard Particle Model spectrum and its properties. Other forms may
include cosmological data, gravitational waves and black hole phenomena.
String theory is a framework which is consistent with perturbative quantum
gravity and enables the calculation of different observable quantities
in a self--consistent framework. In this view, the relevant question
is not whether string theory is the Theory of Everything, but rather
whether any of the ingredients in string theory, {\it e.g.} the appearance
of extra degrees of freedom, in the guise of extra dimensions,
is relevant in the real physical world. 

The fermionic $Z_2\times Z_2$ orbifolds provide a laboratory
to explore how the detailed phenomenology of the
Standard Model and unification emerge from string theory.
Among there are:
  construction of the first Minimal Standard Heterotic--String Model
  (MSHSM) that contains solely the states of the Minimal Supersymmetric
  Standard Model (MSSM) in the effective low energy field theory
  below the string scale \cite{fny, mshsm};
  The prediction of the top quark mass at $\sim 175-180$GeV \cite{tqmp},
  several years prior to its experimental observation \cite{topdiscovery};
  Fermion masses and CKM mixing \cite{fermionmasses};
  Neutrino masses \cite{nmasses};
  Gauge coupling unification \cite{gcu};
  Proton stability \cite{ps};
  Supersymmetry breaking \cite{sd}; 
  Moduli fixing \cite{moduli};
  Since 2003 a detailed classification program of fermionic $Z_2\times
  Z_2$ orbifolds has been carried out
\cite{so10class, so64class, fffclass}
  leading to the discovery of Spinor--Vector Duality (SVD) \cite{svd}
  and exophobic string vacua \cite{so64class}. In recent years,
  the classification program was extended to string vacua with asymmetric
  boundary conditions \cite{asyclass};
  Vacuum stability \cite{vacstab}.
  The vacuum is stable in supersymmetric string
  solutions. Supersymmetry is broken in the real world and its breaking may
  induce vacuum instability in the form of physical tachyonic modes,
  and run away vacuum configurations. Hence, a crucial question is
  whether there exists a stable string vacuum with broken supersymmetry.
  The jury is still out on this question. Moreover, a pertaining question is
  whether the energy in the vacuum is negative or positive. This issue
  has been studied in recent years and is likely to remain an area
  of investigation in the forthcoming years \cite{vacstab}.
  

  In summary, the fermionic $Z_2\times Z_2$ heterotic--string orbifolds
  provide a laboratory to explore how the structure and parameters of the
  Standard Model arise from a fundamental theory of quantum gravity.
  They give rise to a large space of three generation string models that admit
  the canonical $SO(10)$ GUT embedding of the Standard Model spectrum.
  The interesting question is not whether any of these models correspond to
  our physical world, but rather whether any of the structures which are shared
  by these vast space of models is relevant for the structure exhibited
  by the Standard model data, {\it e.g.} in the favour data, which is
  discussed in section \ref{fmasses}. It was argued that the
  phenomenological vitality of the free fermionic models is
  rooted in their underlying $Z_2\times Z_2$ orbifold structure
  \cite{z2xz2}. 

\section{Fermion masses and mixing in string theory}\label{fmasses}

String theory provides the arena for calculating the
Standard Model parameters in terms of the Vacuum Expectation
Values (VEVs) of some moduli fields in the spectrum of
specific string models. The flavour parameters are of particular
interest. The origin of these parameters is
not explained in the context of Grand Unified Theories.
In string theory, the number of generations is related to
the topology of the compactified six dimensional internal
manifold and the flavour parameters are related to its
detailed geometrical properties. The free fermionic models
gives rise to a large number of three generation models
and detailed scenarios of the Standard Model flavour
parameters can be calculated. 

The details of these scenarios are not the crucial point,
and they have been discussed amply before
\cite{tqmp, fermionmasses, nmasses, superpoterm}.
The free fermionic models correspond to $Z_2\times Z_2$ orbifold
compactifications, which have an untwisted sector and
three twisted sectors. These sectors preserve the $SO(10)$ symmetry.
The string models also contain
sectors that arise from from the breaking of the
$SO(10)$ symmetry at the string scale, and give rise to massless
exotic states that do not satisfy the $SO(10)$ quantisation
of any unbroken $U(1)$ symmetry in the Effective Field Theory (EFT)
limit. The Standard Model generations are obtained from the
twisted sectors $b_{1,2,3}$, and in many examples each twisted
sector produces a single generation, thus providing a
plausible explanation for the existence of three
generations in nature \cite{nahe}. The Standard Model Higgs
states are obtained from the untwisted sector as well as from the
twisted sectors.
The details of the Higgs spectrum is model dependent and typically
the existence of at least two Higgs doublets, as required in
$N=1$ supersymmetric models, is imposed, as well as the presence
of a leading Yukawa coupling that can produce a mass term for
the top quark. 
The untwisted sector can produce up to 
three pairs of Higgs doublets, 
$h_i,~{\bar h}_i, i=1,2,3$, and one or two additional pairs, 
$h_{\alpha\beta},~{\bar h}_{\alpha\beta}$, can be obtained from
the twisted sectors.
The scalar and fermion masses 
are obtained from renormalisable and non--renormalisable
terms in the superpotential
\begin{equation}
    {cg}{f_if_j}h{{\left({{\langle\phi\rangle}\over{M}}\right)^{N-3}}}
\end{equation}
where the $c$ coefficients
are calculated from tree level correlators between vertex operators;
$g$ is the gauge coupling at the string scale, 
which is determined by the dilaton VEV;
$f_i$ and $f_j$ are the fermionic states from the twisted sectors
$b_{1,2,3}$; $h$ stands for the light Higgs states in the EFT limit;
$M~\sim~10^{18}~GeV$ is related to the string scale; and
$\langle\phi\rangle$ are Vacuum Expectation Values (VEVs) of
Standard Model singlet fields, along supersymmetric flat directions.
These VEVs are typically of the order of $0.1 M$
producing fermion mass terms
in the low energy EFT that are suppressed compared
to the leading top--quark mass term \cite{tqmp}.
The calculation of the Yukawa couplings for the lighter
fermions is obtained by analysing higher order terms 
in the superpotential and extracting the effective
dimension four operators \cite{superpoterm, fermionmasses}.
Using this methodology, detailed scenarios can be obtained for
the flavour mass and mixing parameters. 
For one specific solution the up and down quark mass 
matrices take the form 
\begin{eqnarray}
& &   M_u\sim\left(
  \begin{matrix}
&    \epsilon &\frac{{V_3{\bar V}_2\Phi_{45}\bar \Phi_3^+}}{{M^4}} &0\cr
&{{V_3{\bar V}_2\Phi_{45}\bar \Phi_2^+}\over{M^4}} 
&{{{\bar\Phi}_i^-\bar \Phi_i^+}\over{M^2}} 
&{V_1{\bar V}_2\Phi_{45}\bar \Phi_2^+}\over{M^4} \cr
&0 &{V_1{\bar V}_2\Phi_{45}{\bar\Phi}_1^+}\over{M^4} &1\cr
\end{matrix}
    \right)v_1,
\label{mu}\\
~~{\rm and}~~
& & 
  M_d\sim\left(
  \begin{matrix}
 &   \epsilon
&{{V_3{\bar V}_2\Phi_{45}}\over{M^3}} &0\cr
&{{V_3{\bar V}_2\Phi_{45}\xi_1}\over{M^4}} 
&{{{\bar\Phi}_2^-\xi_1}\over{M^2}} &{V_1{\bar V}_2\Phi_{45}\xi_i}\over{M^4} \cr
&0 &{V_1{\bar V}_2\Phi_{45}\xi_i}\over{M^4} 
      &{{\Phi_1^+\xi_2}\over{M^2}}\cr
  \end{matrix}
  \right)v_2,
\label{md}
\end{eqnarray}
The down and up quark mass matrices are diagonalized 
by bi--unitary transformations
$$
  U_LM_uU_R^\dagger=D_u\equiv{\rm diag}(m_u,m_c,m_t),~~~~~
  D_LM_dD_R^\dagger=D_d\equiv{\rm diag}(m_d,m_s,m_b),
$$

with the CKM mixing matrix given by 
$$V=U_LD_L^\dagger.$$ 
The VEVs of $\xi_1$ and $\xi_2$ are fixed to be $\langle \xi_1 \rangle 
\sim M/12$, and $\langle \xi_2 \rangle \sim M/4$, by the masses $m_s$ and $m_b$
respectively.
Substituting the 
VEVs and diagonalizing $M_u$ and $M_d$ by a bi--unitary transformation, we 
obtain the mixing matrix
\beq
\vert V \vert\sim \left(
\begin{matrix}
  0.98&0.205&0.002 \cr 
  0.205&0.98&0.012 \cr 
  0.0004&0.012&0.99 \cr
\end{matrix}  \right).
\label{ckm}
\eeq
It is important to stress that the mass matrices in (\ref{md}) are
for illustrative purposes only and that our understanding
of string theory is not at a stage where we can reliably predict the
flavour parameters directly from the theory. At best we can
calculate the Yukawa couplings in terms of the gauge coupling at the
string unification scale, which is fixed by the dilaton VEV. The dilaton VEV
is undetermined at present and our understanding of the dynamics that may
fix it are very rudimentary. 
A rather robust prediction in the free
fermionic heterotic string models is $\lambda_{\rm top} = \sqrt{2} g$,
where $g$ is the gauge coupling \cite{tqmp}. The Yukawa couplings for the
lighter quarks and leptons vary between models.
It is seen from eqs. (\ref{md}, \ref{ckm}) that viable mass matrices may in
principle be obtained. 

The crucial fact, however, is that in all of these scenarios,
as well as in many other string models on the market,
the fermion masses arise from a coupling between fundamental
fermionic states in the string spectrum to a fundamental
scalar state in the string spectrum, which is identified with
the Standard Model Higgs field. The nature
of the Higgs boson therefore become the crucial question from the
point of view of the string models. Is the Higgs a fundamental
or composite particle? In the case that the Higgs is a composite
particle, the entire edifice of $SO(10)$ inspired string model
building can be discarded as irrelevant to our observed world.
It is important to emphasise that observation of a composite
Higgs will not necessarily nullify the relevance of string theory
as a scheme for synthesising the Standard Model with gravity.
It will show that a large class of models, and in fact
an entire direction of research, is simply off track. 
A different approach to string model building will
have to be thought if the Higgs is composite.
It is further important to reflect that the question of compositeness
is inherent in string theory. Elementary particles in string
theory have an internal structure. The string bits approach to string
theory testifies to that view \cite{stringbits}.
The question might be regarded as whether the internal structure sets
in at a scale close to the Electroweak symmetry breaking mechanism
or, alternatively, whether it sets in at a high scale, close
to the Planck scale.

\section{The Higgs Enigma }

Whether the Higgs particle is fundamental or composite is
a question that Future Collider Facilities will be able to probe.
In the very least, they will set limits on its scale of
compositeness. The study of the Higgs boson is a well defined
problem from the experimental point of view. The Higgs is there.
We know where it is. If took decades to find it. At least since the
early 70s and the demonstration of renormalisability of spontaneously
broken non--Abelian gauge theories. Exploring
the properties of the Higgs boson will advance our
fundamental understanding of nature, with substantial
capacity for accolade and prizes from its experimental
exploration. The physics of the Higgs boson and Electroweak
symmetry breaking is often associated with new sectors
Beyond the Standard Model (BSM). Whether nature takes advantage of
any of these new realms, or perhaps a yet unforeseen territory,
it will be the task of the experimentalists to navigate the
new frontiers.

The hierarchy problem motivated many of its extensions. Supersymmetry
extends the symmetry of the SM to include symmetries under exchange
of fermion and bosons. Supersymmetry is a well motivated theoretical
framework. The spacetime Poincare symmetry and the internal
symmetries of the gauge interactions are the bedrock of the SM.
It is natural to extend these symmetries to Fermi--Bose symmetry. 
The relatively light Higgs mass is also compatible with the supersymmetric
extension of the SM. Furthermore, the top quark mass and Yukawa coupling
are instrumental in inducing radiative Electroweak symmetry breaking in
supersymmetric Grand Unified Theories. 
In its local version, supersymmetry includes gravity,
and supersymmetry is realised naturally in string theory.
In this framework, the Higgs particle is a fundamental scalar
state, similar to the other fermionic and bosonic particles
in the spectra of these theories. 
While this are well motivated mathematical constructions,
especially in the context of $SO(10)$ string GUTs, nature
may have chosen a different route.

An alternative explanation to the hierarchy problem is
that the Higgs particle is not fundamental, but is a composite
of yet more fundamental constituents. Inspiration from
manifestation of symmetry breaking in other physical
systems would in fact suggest that this is the route
chosen by nature, albeit we have no clue what the fundamental
constituents may be. 

Composite Higgs models
(see {\it e.g.} \cite{Contino:2010rs} for a review),
used to solve the hierarchy problem, are regarded as
natural if the hyper--pion decay constant $f_\Pi$ is
below $\mathcal{O}(\mathrm{TeV}).$
Such models imply a plethora of resonances,
but their couplings and spectra are model dependent.
A new discovery collider would be required to produce them.
Unless we can directly study Higgs scattering at energies above
$f_\Pi,$ the smoking gun for composite models would be
deviations from  the Standard Model Higgs couplings. 

A relatively model--independent prediction is that the electroweak vector
boson mass terms are proportional to $\sin^2 \left(\frac{v + h}{f_\Pi}\right)$
where $v$ is the expectation value of the Higgs field
(which is not exactly equal to the value obtained from the
electroweak fit $v_{\rm EW} \simeq 246$ GeV).
This structure entails that the couplings between the
the massive vector bosons and Higgs boson
are modified by a factor $\cos v/f_\Pi$ compared to the
Standard Model values,
yielding deviations of order
$(v/f_\Pi)^2.$ Estimates for the HL--LHC will limit these
parameters (interpreted as $\kappa_W, \kappa_Z$) to be
within $1\%$ of the Standard Model value \cite{deBlas:2019rxi},
producing a lower limit on $f_\Pi \gtrsim 2$ TeV
Future Lepton or Hadron collider will constrain them to be
within $0.2\%$ of the Standard Model values,
or $f_\Pi \gtrsim 6$ TeV.

Deviations of the vector boson couplings are not special to composite
Higgs models. It will also be necessary to confirm any deviation by a 
measurement of the {\it\bf triple--Higgs--coupling.} The deviations
of this parameter from the Standard Model value are more difficult to
compute, and less model--independent.
A calculation based on the potential  \cite{Contino:2010rs}
$$V(h) \simeq \alpha \cos\frac{v+h}{f_\Pi} - \beta \sin^2 \frac{v+h}{f_\Pi} 
$$
gives a deviation of the Standard Model triple Higgs coupling
$\kappa_3 = \cos v/f_\Pi$. The HL--LHC is only expected to have a
sensitivity of order $50\%$ for this measurement. A new high--energy
collider is required. The FCC--ee will not have sufficient energy.
A new hadron collider with 50--60 TeV CoM energy might be able to
reach a precision of a few percent.
Combined with the above measurements, it will be able to refute or confirm
whether compositeness is relevant for the electroweak symmetry breaking
mechanism.

The question of whether the Higgs particle is composite or fundamental
is therefore of vital importance for physics Beyond the Standard Model,
including for its realisation in string constructions. The
experimental interest in Higgs physics is of course much wider and the
task of future experiments will be to measure the Higgs couplings
to better and better precision, among them its couplings to the
Standard Model fermions and bosons. The cubic and quartic Higgs
self--couplings are of pivotal interest in elucidating the
electroweak symmetry breaking mechanism. Much work is of course
going to the theoretical analysis of experimental measurements
of the Standard Model and Beyond the Standard Model parameters.
Of special note are perhaps the efforts into the Standard Model
Effective field Theory analysis, which continue to develop
the approach of \cite{hagiwarazeppenfeld}, and in essence that of
\cite{fermi}. 

\section{Beyond the Standard Model}

The Standard Model of particle physics is compatible with all
subatomic observational
data to date. There are compelling arguments that it is an effective
field theory, which is only valid up to some cutoff scale. The
naturalness argument suggests that the cutoff scale is of the order
of the TeV scale, which will be investigated in Future Collider Facilities
(FCF), though naturalness arguments may only reflect
a lack of a deeper understanding of the mathematical description. 

\begin{figure}[h]
\includegraphics[width=16pc]{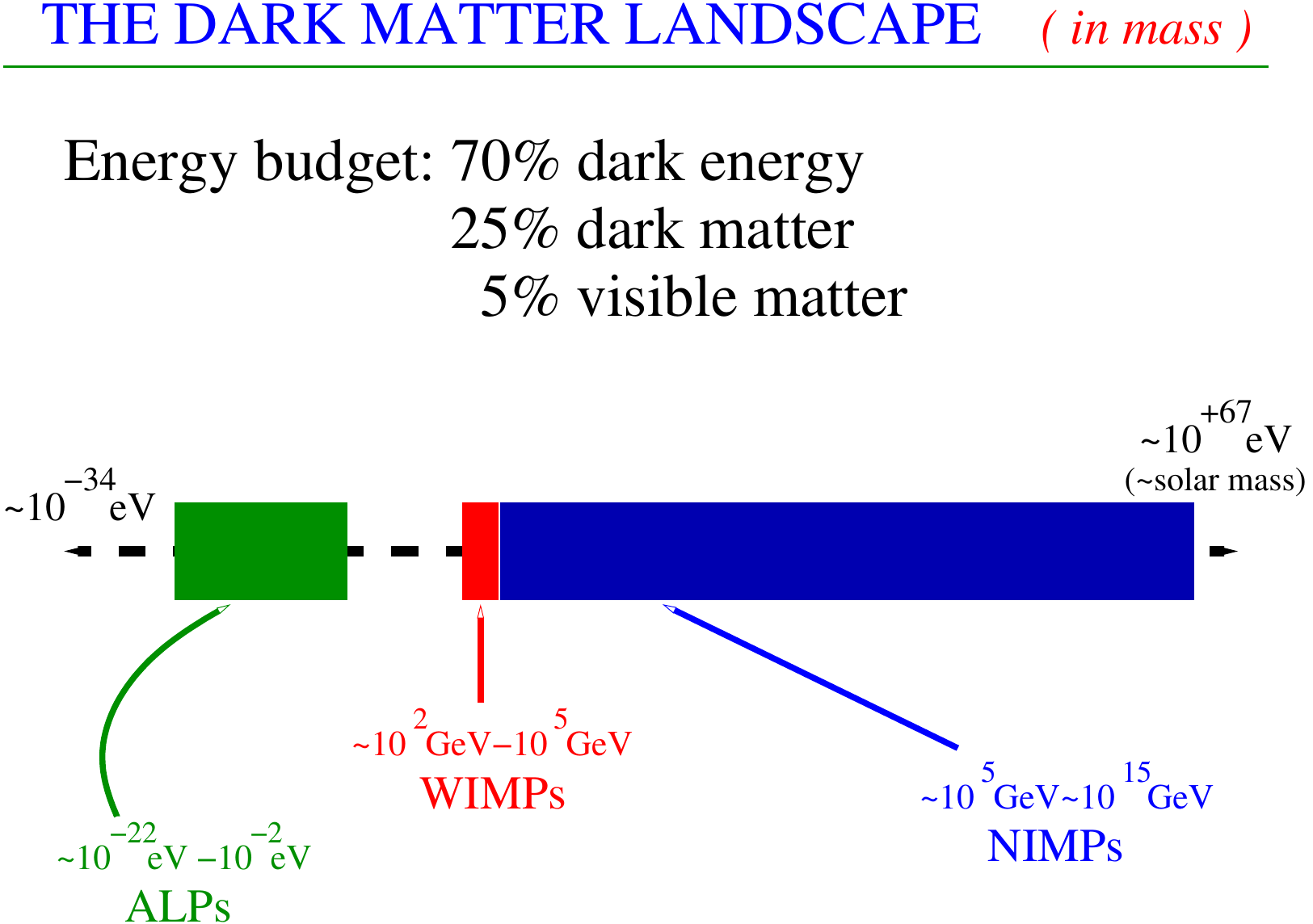}\hspace{2pc}%
\begin{minipage}[b]{12pc}
\caption{\label{darkmatters}%
\emph{
Dark matter candidates vary from ultra--massive to ultra--light,
spanning some eighty orders of magnitude in mass scale.
While the evidence for dark matter is compelling,
the properties of specific dark matter candidates are
not well motivated from a theoretical point of view and
not well defined from an experimental point of view. 
}
}
\end{minipage}
\end{figure}

The fundamental dichotomy between quantum mechanics, in its manifestation
as the Standard Model of particle physics, and gravity leads to a few
inconsistencies between the observational data and the mathematical modelling,
that suggest the existence of physics Beyond the Standard Model.
Gravitational based data suggests the existence of dark matter and dark
energy, which make up 95\% of the energy budget in the universe, whereas
the visible matter only accounts for 5\%. However, the properties of
specific dark matter candidates are not well defined from an experimental
point of view and are not particularly well motivated from a theoretical
point of view. Dark matter candidates can have varied properties, including:
Weakly interacting Dark Matter; Self-interacting dark matter; Super-massive
dark matter; axions; ultra-light dark matter; primordial black holes; ...
The bazaar of possibilities is depicted qualitatively in Figure 
\ref{darkmatters}.

None of the dark matter candidates is particularly well motivated from a
theoretical point of view. If we consider, for example, the Weakly Interacting
Massive Particle (WIMP) in the context of the supersymmetric
extensions of the Standard Model (SM). Supersymmetry is reasonably well
motivated and alleviates the hierarchy problem. A dark matter candidate is
an added bonus. However, in order for the lightest supersymmetric particle
to be stable, one has to assume the existence of $R$--parity to avoid
problems with rapid proton decay, which is ad hoc and in general expected to
be violated in string theory and other approaches to quantum gravity.
Axion candidates do not seem to fare better in this respect.

We can look in string theory for a guide to promising scenarios of
physics Beyond the Standard Model. String theory is a rather constrained
approach to quantum gravity. It predicts the existence of a fixed number
of additional degrees of freedom, often guised in the form of extra
dimensions, that are needed for the consistency of the theory.
The first task of a new theory is to reproduce the physics
that we are familiar with. String theory produces models that
reproduce the gross structure of the Standard Model and its Minimal
Supersymmetric extension, as well as its embedding in $SO(10)$
Grand Unification. We can regard such models as bench mark models
for BSM physics. As discussed above, the fermionic $Z_2\times Z_2$
orbifold produces numerous three generation models that admit the
$SO(10)$ embedding of the Standard Model chiral generations.
These models contain a number of BSM $U(1)$ symmetries. Some of these
$U(1)$ symmetries are in the observable sector, {\it i.e.} the
Standard Model states are charged under them, and may be family
universal or non--universal. Some of the extra Abelian and non--Abelian
symmetries may arise in an hidden sector in the sense that the
Standard Model states are not charged under them. Generating viable
fermion masses implies that the majority of
the extra $U(1)$ symmetries have to be broken near the string
scale \cite{fermionmasses}. It was argued that family universal
$U(1)$ symmetries at low scales may be instrumental in suppressing proton decay
mediating operators and the supersymmetric $\mu$ parameter
\cite{pstabzp}. However, compatibility with the low scale
gauge parameters requires that an extra low scale $Z^\prime$
as an embedding in $E_6$ \cite{pstabzp}, whereas some of the
extra $U(1)$ symmetries are not embedded in $E_6$. Extra string inspired
$E_6$
$Z^\prime$ vector bosons have been studied intensely since the mid--eighties
as possible signatures of string theory \cite{zpreports}.

A string derived $Z^\prime$ model with $E_6$ embedding proved, however,
more elusive to obtain. The reason is that $E_6$ is broken
at the string level and a remnant family--universal $U(1)$
symmetry becomes anomalous at the string scale \cite{auone}
and cannot remain unbroken down to low scales. In \cite{frzprime}
we used self--duality under the Spinor--Vector duality \cite{svd}
to produce a string model that, while the $E_6$ gauge symmetry
is broken, the chiral generations complete full $E_6$
multiplets. In this case the extra $E_6$ $U(1)$ is anomaly free
by virtue of its embedding in $E_6$ chiral representations.
An alternative to this construction is to embed the extra
$U(1)$ in a non--Abelian group factor as in the
$SU(6)\times SU(2)$ string GUT model \cite{su62}.
Because of the restricted heavy Higgs states in the string models,
the string derived $Z^\prime$ model allows for a unique
combination of $U(1)$ symmetries to remain unbroken down to
low energies that gives rise to experimental signatures
beyond the Standard Model \cite{zppheno, McFG, wdm}.
The motivation for the extra $U(1)$ symmetry to remain
unbroken down to the TeV scale stems from the solution that
it provides for the suppression of the supersymmetric $\mu$--parameter.
The Minimal Supersymmetric Standard Model (MSSM) have two Higgs
doublets in vector--like representation and their mass scale is
controlled by the $\mu$--parameter. There is no symmetry
that protects the $\mu$ parameter. It is set by hand to be of the
order of the electroweak scale to produce a light Higgs eigenstate
in agreement with the one observed at the LHC. In the $Z^\prime$ model
the Higgs doublets are chiral under the extra $U(1)$ symmetry
and the $\mu$--parameter is generated by its breaking.
While the existence of the symmetry, and the associated
massive $Z^\prime$ vector boson, is reasonably well motivated,
there is no guarantee that it lies within reach of
forthcoming collider experiment. Even at the FCC--hh with 100TeV
Centre of Mass (CoM) energy. It would seem folly 
to design an experiment with no guarantee of producing substantial
gain in our understanding of nature, in the form of improved
measurements of the prevailing mathematical formalism,
but solely to hinge it on constraining the parameter space
of a rather hypothetical theoretical proposition.
On the other hand, the possibility of making substantial new discoveries
mandates that we explore the new energy frontiers.

As another example, I consider the ambiguity in pinning down
the particle properties of string derived dark matter candidates.
The string derived models give rise to exotic matter states that
carry fractional charge under any of the unbroken GUT $U(1)$ symmetry. 
One class of such states carry fractional electric charges
and must be stable due to electric charge conservation.
They must be sufficiently massive and/or sufficiently rare
to avoid the experimental constraints.
Such states must appear in modular invariant string models that
possess the canonical GUT embedding of the weak hypercharge
\cite{bert}. However, these fractionally charged states
can appear solely in the massive string spectrum, and not
necessarily as massless states \cite{so64class}.
Alternatively, they must be non--chiral and become massive
at a high mass scale \cite{fc}.

In addition to the exotic states that carry fractional electric charges,
the string Standard--like
Models can contain states that carry the standard charges under
the Standard Model (SM) gauge group, {\it e.g.} are SM singlets, but carry
fractional exotic charges under another observable $U(1)_{Z^\prime}$
in the string models.This extra $U(1)$ is necessarily broken.
Breaking this extra $U(1)_{Z^\prime}$ by states that carry the standard GUT
charges leaves a discrete symmetry that forbids the decay of the
exotic states to the SM states \cite{SSR, wdm}. These exotic states
provide viable dark matter candidates \cite{SSR, wdm}.
However, also in this rather well defined scenario, their
properties as dark matter candidates vary widely and
depend on the $U(1)_{Z^\prime}$ breaking scale and whether
inflation is assumed or not. The different possibilities are
enumerated below:
\begin{enumerate}\label{diffdmpossi}
\item $M\gg M_{Z^\prime}$ without inflation~~~~~~~~~
  {$\Rightarrow M\le 10^{5}~\hbox{GeV}$}
\item { $M\gg M_{Z^\prime}$ with inflation and $T_R > M_{Z^\prime}$~~~~~
{$\Rightarrow M > T_R\left[25+{{1}\over{2}}\ln\left({{M}\over{T_R}}\right)\right].$}}
\item  { $M\ll M_{Z^\prime}$ without inflation ~~~~~~~
{$\Rightarrow M<3~\hbox{keV}$}}
\item { $M\ll M_{Z^\prime} ~~\hbox{with inflation}~~
M~~
\begin{cases}
>~T_R\left[25+{{1}\over{2}}\ln\left({{M^5}\over{M_{Z^\prime}^4T_R}}\right)\right]~,
                 &   T_R < M \cr%
<~{{M^4_{Z^\prime}\over{T_R^3}}6.9\times 10^{-25}%
\left({{g_*}\over{200}}\right)^{1.5}%
{{1}\over{N_{Z^\prime} g^2_{\rm eff}}}},& T_R > M
\end{cases}
$}
\end{enumerate}
where
$M_{Z^\prime}$ is the mass of the $Z^\prime$ vector boson
and $M$ is the dark matter mass. 
This list shows that even in this rather restricted model,
the properties of the dark matter are not pinned down.
Additionally, hidden sector glueballs have been proposed
as Self Interacting Dark Matter candidates \cite{fp}.
While the gravitational based observations provide
compelling motivation for the existence of dark matter,
its properties are not well defined from an experimental
particle physics point of view. 
It would seem prudent if experiments that are being planned with that
objective in mind, identify the potential added benefit in terms
of constraints on the prevailing parameterisation of observational data,
{\it i.e.} the Standard Model parameters, including its neutrino sector.
Proton decay experiments that proved instrumental as neutrino detectors
can serve as past examples.  

\section{Future Collider Facilities}

The discovery of the Higgs boson concluded a glorious century of experimental
discoveries, starting with the discovery of the nucleus in 1911, and
paves the way for future experimental studies and potential discoveries.
The Higgs boson is there for the picking.
The future of the experimental particle physics
program has been under community discussion over the past few years
\cite{Butler:2023eah, Gourlay:2022odf}. 
Careful consideration is required, 
given the level of public resources that are needed 
to pursue the experimental programs.
In this context we have to weigh the potential science benefit
as well as the economic and social. 
Priorities have to be established in the face of finite resources
and constraints emanating from wider public needs.
We participated in this discourse in several past publications
\cite{McFG, thequest, fgg}. 
As discussed above the string derived $Z^\prime$ model provides
a bench mark model. I recap some of this arguments briefly here.

Historically, the modern scientific method is a European development,
starting with the renaissance, Galileo and Newton are especially noted
for developing the mathematical modelling of empirical observations.
The age of enlightenment brought about an explosion of scientific
breakthroughs that established the mathematical formalism of classical
mechanics, thermodynamics and electromagnetism, providing the
concepts, {\it e.g.} the Lagrangian and the Hamiltonian, that are
still used today. Rutherford peered into the nucleus and
opened the nuclear age. With the 20th century European upheavals,
the leadership transferred to the US, which led the field of experimental
particle physics from the late 30s to the 70s. CERN was founded by the
European community in the 1950s to compete with the US.
With the demise of the SSC in 1993 the leadership reverted back
to Europe. At present CERN is the world centre for experimental
particle physics research. Its role goes beyond its scientific
mission. It serves as a platform for collaboration and cooperation
in humanity's never ending quest to understand the world we live in.
Today, one could argue, the technical know how in experimental particle
physics research resides solely in Europe and CERN, but it exists in principle
in the US, China, Japan and Russia, though not at the same level of practice.

The science questions are varied and are all extremely interesting.
I argued above that priority should be given to experimental
studies of the Higgs boson. The next questions are the how, when
and where? Which scientific instruments should be built? their timeline
and where should they be built.

As the leading world laboratory in experimental particle physics,
CERN is seeking to follow its LEP playbook. The HL--LHC will run
until the late 2030s--early 2040s. Concurrent with the HL--LHC,
a 91km circular tunnel will be dug to accommodate the Future
Circular Collider. Following the LEP playbook, the first
phase will be a lepton machine that uses warm, and hence cheaper,
magnets. The FCC--ee will operate up to $\sim$350GeV and will do
primarily bread\&butter precision measurements of the Standard Model
parameters. It will have limited capacity to explore Beyond the
Standard Model physics and limited capacity to probe the Higgs triple and
quartic couplings as well as the top Yukawa coupling. The FCC--ee will be
followed by the FCC--hh in the same tunnel. The vision is to use new
superconducting technology, based on Niobium--Tin (${\rm Nb}_3{\rm Sn}$)
alloys, that will enable construction of
magnets with 16 Tesla magnetic field. The current LHC magnets use
Niobium–Titanium (NbTi) alloys. They operate at
temperature of 1.9K and generate a magnetic field of 8.3
Tesla. The design and manufacturing of the magnets is
the key elements in both the cost and success
of the collider experiments.  
An insightful account of the role of the magnets in the
SSC story is given in \cite{tunnelvisions}.
It is important to understand the scale of the challenge.
What is required in an accelerator project is not merely a
prototype that works. What is needed, is a sufficiently robust design
that can be manufactured on an industrial scale to build a $\sim$91km ring of
magnets that will produce a stable magnetic field over a period
of $\sim$30 years. An enormous challenge. The FCC--hh is planned to
operate at 100TeV CoM energy starting in the 2070s.

China and the US are also in advanced planning stages of
their experimental particle physics future. At the time of writing China
put on hold its plans which parallels the European plan of an initial 
lepton collider to be followed by a hadron collider.
Absence of an on--going active experiment may give the
Chinese a slight advantage over the potential timeline.
The recommendation of the P5 committee in the US is
the construction of a multi--TeV muon collider,
which represents a major technological challenge. The immediate
interest here is the development of the accelerator
technology. Given the experience of the MICE experiment 
\cite{Gregoire:2003nh, MICE:2019jkl}, it is perhaps reasonable to
expect that a muon collider may be feasible sometime in the future. 

The International Linear Collider (ILC),
which is a linear $e^+e^-$ collider,
is  another major item on the planning boards. 
The ILC can reach higher CoM energies, 
possibly up to 1TeV, which
will enable studying the Higgs sector at higher energies
compared to the FCC--$e^+e^-$ and more precisely relative to a circular
hadron collider. Its drawback is the reduced luminosity compared to a
a circular lepton collider.
The ILC will operate at different stages, {\it i.e.}
at 90 GeV, 250 GeV, 350 GeV, 500 GeV, and 1 TeV.
With the time intervals between its different phases,
it will be a prolonged project over many decades.
Japan is a potential host of this project, though at present there is
no clear commitment. It would also seem prudent to first identify
its clear physics targets beyond those of the FCC--ee. The SLC experience,
which operated in parallel to LEP may serve as a guide. The advantage of
the ILC is that it can reach higher energies and its construction is
incremental. Its disadvantage is the reduced luminosity.

An alternative route that we proposed over the past few years
is the construction of an Upgraded Superconducting Super Collider
(USSC) \cite{McFG, thequest, fgg}. The core element of this proposal
is to utilise
existing superconducting alloy technology in an FCC--size tunnel,
which will allow CoM energies of the order of 50--60 TeV, {\it i.e.}
up to four times the LHC CoM energy. The basic parameters of the proposed
collider are near those of the SSC, hence its dubbing as an
Upgraded Superconducting Super Collider. The Original SSC magnet
were designed with 6.6 Tesla at 4.5K and 5cm bore producing a 40TeV
CoM collisions.
One can envision using upgraded magnet technology, {\it e.g.} the
LHC magnets with 8.3 Tesla at 1.9K and 56mm bore, will optimally reach
~14TeV Com energy in its 27km ring. Achieving four times the LHC Com energy
is feasible. This proposal can be termed as Upgraded SSC collider (FCC--LHC),
or as FCC--LHC, {\it i.e.} FCC--like tunnel but with available
magnet technology. 

In terms of its physics reach, the USSC will be able to do Bread\&Butter
Standard Model physics. It will improve the measurements of the SM parameters
including the triple and quartic Higgs couplings as well as the top Yukawa
coupling. Additionally, it will have a fair shot at discovering new physics,
provided that it exists within the 5--6 TeV region. In ref. \cite{McFG, fgg}
we demonstrated that the cross section for $Z^\prime$ production with 50TeV
CoM energy increases by
three orders of magnitude compared to the LHC at 14TeV. This increase
will be across the board, demonstrating the capacity of the USSC for SM and
BSM processes. 

The physics case for the USSC/FCC--LHC is strong. The Original SSC (OSSC)
was supposed to
start operations in the 1996--2000 region. The site was chosen in October
1988. Given this timeline, it is reasonable to expect that the USSC
construction can be completed in 10--15 years from decision, {\it i.e.} the
USSC could become operational in the late 2030s -- early 2040s.
Hence contributing substantially to the sustainability of the field,
and preservation of the expertise. 

The SSC price tag at the time of its cancellation in October 1993 was
\$11B, and accounting for inflation, the cost of the USSC today would be
\$25B. This is a rough estimate as some costs have gone up, whereas
others have gone down. In particular, the cost of the magnets should be
substantially cheaper, given that the expertise are much more prevalent and
the proposal aims to use well established technology.

The fact that the USSC/FCC--LHC is not on the planning board of any of the
major contemporary players, gives the opportunity to new players,
that do not currently possess the technical know-how,
to enter the field.
As the required magnet technology is off the shelf technology,
we proposed in \cite{McFG, thequest, fgg}
that the USSC can be pursued as a Middle East project and
funded by Saudi-Arabia and other regional countries.
The synchrotron light source at the SESAME site in Jordan
is an established regional collaborative effort and can
serve as the seed for the USSC. Following CERN
which is located in a relatively neutral site, the SESAME
site in Jordan can play a similar role for the USSC/FCC--LHC.
The potential reward in terms of technological development
and international prestige is substantial.
The establishment of the USSC/FCC--LHC with successful physics delivery
will be an enormous achievement.
Following the CERN experience of providing a platform for collaboration,
the USSC/FCC--LHC can serve as a Project for Peace, 
promoting curiosity driven partnership between neighbouring countries.
The discovery of new physics at the USSC/FCC--LHC will not only transform
the field of particle physics, but will raise the regional sponsors into the
world leaders in humanity's quest to understand the universe we live
in and the rules that govern it. 

\section{Conclusions}

The discovery of the Higgs boson by the ATLAS and CMS collaborations
of the LHC experiment at CERN concludes the glorious century of
experimental discoveries. Starting with Rutherford's discovery
of the Nucleus in 1911, the discoveries of quarks and leptons
in the 1950s to 1970s, and the discoveries of weak
vector bosons in the 1980s, the Higgs boson discovery
cemented the Standard Model of particle physics as
providing the viable parameterisation of all observable subatomic
phenomena up to the TeV scale, and possibly up to much
higher scales. The Higgs discovery opens the road to
the experimental exploration of years to come.
Elucidating the Higgs properties and the parameters
that govern its self--interactions and those
with the other bosons and fermions is the top priority
of experimental particle physics research. These searches
may also reveal the existence of new particles that
have not yet made their appearance in the laboratory. From the point of view
of Grand Unified Theories and heterotic--string GUTs the
question of whether the Higgs is composite or fundamental
is of prime interest. In these theories the Higgs is
a fundamental state in the spectrum and its coupling
to the Standard Model fermions enables the development
of detailed flavour scenarios. Should the Higgs be composite,
it will indicate that this is not the route chosen by nature.
Should that be the case, 
perhaps a new strongly interacting sector is required,
possibly emerging from theories like those of
\cite{terazawa, RS, mark}.

\section*{Acknowledgements}

I would like to thank Mark Goodsell and Marco Guzzi for collaboration
and discussion. This work is supported in part by
a Royal Society Exchange grant IES/R1/221199, ``Phenomenological studies
of string vacua", and by the STFC Consolidated Grant ST/X000699/1.

\end{document}